\documentclass{aa}
\usepackage{graphicx}
\usepackage[varg]{txfonts}
\usepackage{longtable}
\usepackage{hyperref}
\defcitealias{Liuzzo2013}{Paper~I}

\newcommand{\paperi}{{\citetalias{Liuzzo2013}}}
\newcommand\pflux{\mbox{${\rm \, ph \,\, cm^{-2} \, s^{-1}}$}}

\begin{document}

\title{Exploring the bulk of the BL Lac object population.}
\subtitle{II. Gamma-ray properties.}

\author{F. D'Ammando\inst{1}\fnmsep\thanks{Email: dammando@ira.inaf.it} \and M. Giroletti\inst{1} \and S. Rain\`o\inst{2,3}}

\institute{INAF Istituto di Radioastronomia, via Gobetti 101, I-40129 Bologna, Italy
\and Dipartimento Interateneo di Fisica "M. Merlin" dell'Universit\`a e del Politecnico di Bari, I-70126 Bari, Italy
\and Istituto Nazionale di Fisica Nucleare, Sezione di Bari, I-70126 Bari, Italy}

\date{Received ; accepted }

\abstract
{}
{We are studying an unbiased sample of 42 nearby ($z<0.2$) BL Lacertae objects with a multi-wavelength approach. The results of Very Long Baseline Interferometry observations were presented in the first paper of this series. In this paper, we study the $\gamma$-ray properties of the sample.}
{We analyse data collected by the {\em Fermi} Large Area Telescope (LAT) during its first 8.5 years of operation in the energy range $0.1-300$ GeV.}  
{We reveal 23 sources with a test statistic greater than 25 (corresponding to $\sim$ 4.6-$\sigma$) out of 42, with 3 sources not detected in the third LAT active galactic nucleus (AGN) catalogue, and fluxes between $3.5\times 10^{-10}$ and $7.4\times10^{-8}$ \pflux. The majority of the sources have hard spectra ($\Gamma \leq 2$), with only four having values in the range 2.1--2.4. The three newly detected sources have fluxes in the range between $0.54\times10^{-9}$ and $1.35\times10^{-9}$ \pflux\, and photon index 1.7--1.9. Among the 23 LAT-detected sources, 19 are included in the third catalogue of hard {\em Fermi}-LAT sources, with a spectrum that connects relatively smoothly from 0.1 GeV to 2 TeV. LAT-detected BL Lacs are more luminous on parsec scales with respect to non-LAT-detected sources and have larger core dominance according to the unified models.} 
{The LAT-detected BL Lacs seem to be composed of a bulk of "classical" sources dominated by Doppler boosting and characterised by compact and bright radio emission as well as hard $\gamma$-ray spectra. Moreover, we have identified a possible population of low-luminosity BL Lacs not detected by LAT, lacking a VLBI core, and with a small Doppler factor. Furthermore, three LAT-detected sources show non-classical properties for $\gamma$-ray emitting BL Lacs (no evidence of relativistic jet, low Doppler factor in radio images, relatively low core dominance) and three other sources, while showing radio emission on parsec scales, are not detected in $\gamma$ rays so far.} 

  
\authorrunning{F.\ D'Ammando, M.\ Giroletti, \& S.\ Rain\`o}

\maketitle
%

\section{Introduction}

Blazars are the rarest and yet most extreme class of active galactic nuclei (AGNs), characterised by the presence of a relativistically beamed jet of plasma pointing within a few degrees of our line of sight. On the basis of their optical spectra, they are classified as flat spectrum radio quasars (FSRQs, characterised by strong emission lines) or BL Lacertae objects (BL Lacs, with nearly featureless optical spectra). Besides the difference in optical spectra, the two classes differ in many other observational properties, including luminosity, accretion regime, spectral energy distribution (SED), morphology and kinematics of the radio jet, and evolution over cosmic time \citep[e.g.][]{ref:3LAC, ghisellini17, jorstad17}.

\begin{table*}
\caption{List of BL Lac sources in our sample. Sources are sorted by increasing RA.}
\label{tab:table1}
\centering
\begin{tabular}{ccccccc}
\hline\hline
Source Name  & RA (J2000) & Dec (J2000) & Redshift  &  3FGL Name  & Photon Index & Flux (E $>$ 0.1 GeV) \\
             & deg & deg  &           &             &              & (10$^{-9}$ \pflux) \\
\hline
J0751+1730 & 117.8545 & 17.5142 & 0.185 & $-$ & & \\
J0751+2913 & 117.7899 & 29.2265 & 0.194 & $-$ & & \\
J0753+2921 & 118.3525 & 29.3589 & 0.161 & $-$ & & \\
J0754+3910 & 118.6545 & 39.1799 & 0.096 & $-$ & & \\
J0809+3455 & 122.4120 & 34.9270 & 0.083 & 3FGL\,J0809.6+3456 & 1.67 $\pm$ 0.13 & 1.56  $\pm$ 0.60 \\
J0809+5218 & 122.4548 & 52.3163 & 0.138 & 3FGL\,J0809.8+5218 & 1.88 $\pm$ 0.02 & 34.90 $\pm$ 1.83 \\  
J0810+4911 & 122.7275 & 49.1844 & 0.115 & $-$ & & \\
J0847+1133 & 131.8039 & 11.5639 & 0.199 & 3FGL\,J0847.1+1134 & 1.74 $\pm$ 0.12 & 2.86 $\pm$ 0.95 \\
J0850+3455 & 132.6508 & 34.9230 & 0.145 & 3FGL\,J0850.2+3500 & 1.92 $\pm$ 0.20 & 2.09 $\pm$ 1.10 \\
J0903+4055 & 135.8113 & 40.9333 & 0.188 & $-$ & & \\
J0916+5238 & 139.2164 & 52.6412 & 0.19  & $-$ & &  \\
J0930+4950 & 142.6565 & 49.8404 & 0.187 & 3FGL\,J0930.0+4951 & 1.45 $\pm$ 0.21 & 0.56 $\pm$ 0.40 \\
J1012+3932 & 153.2432 & 39.5442 & 0.171 & $-$ & & \\
J1022+5124 & 155.5526 & 51.4001 & 0.142 & $-$ & & \\
J1053+4929 & 163.4339 & 49.4989 & 0.14  & 3FGL\,J1053.7+4929 & 1.80 $\pm$ 0.10 & 3.63 $\pm$ 0.93 \\
J1058+5628 & 164.6572 & 56.4698 & 0.143 & 3FGL\,J1058.6+5627 & 1.95 $\pm$ 0.03 & 32.40 $\pm$ 1.62 \\
J1120+4212 & 170.2003 & 42.2034 & 0.124 & 3FGL\,J1120.8+4212 & 1.62 $\pm$ 0.06 & 5.04 $\pm$ 0.79 \\
J1136+6737 & 174.1253 & 67.6179 & 0.136 & 3FGL\,J1136.6+6736 & 1.72 $\pm$ 0.08 & 3.21 $\pm$ 0.73 \\
J1145$-$0340 & 176.3963 & $-$3.6671 & 0.167 & $-$ & & \\
J1156+4238 & 179.1940 & 42.6354 & 0.172 & $-$ & & \\
J1201$-$0007 & 180.2758 & $-$0.1171 & 0.165 & $-$ & & \\
J1201$-$0011 & 180.4319 & $-$0.1872 & 0.164 & $-$ & & \\
J1215+0732 & 183.7958 & 7.5346 & 0.136 & $-$ & & \\
J1217+3007 & 184.4670 & 30.1168 & 0.13 & 3FGL\,J1217.8+3007  & 1.97 $\pm$ 0.02  & 57.70 $\pm$ 2.75 \\
J1221+3010 & 185.3414 & 30.1770 & 0.182 & 3FGL\,J1221.3+3010 & 1.97 $\pm$ 0.02  & 15.70 $\pm$ 1.67 \\
J1221+2813 & 185.3820 & 28.2329 & 0.102 & 3FGL\,J1221.4+2814 & 2.10 $\pm$ 0.03  & 52.50 $\pm$ 2.65 \\
J1221+0821 & 185.3836 & 8.3623 & 0.132 & $-$ & & \\
J1231+6414 & 187.8808 & 64.2384 & 0.163 & 3FGL\,J1231.5+6414 & 1.94 $\pm$ 0.20  & 2.11 $\pm$ 1.18 \\
J1253+0326 & 193.4458 & 3.4417 & 0.066 & 3FGL\,J1253.7+0327  & 1.84 $\pm$ 0.10  & 5.30 $\pm$ 1.45 \\
J1257+2412 & 194.3830 & 24.2111 & 0.141 & $-$ & & \\
J1341+3959 & 205.2713 & 39.9959 & 0.172 & 3FGL\,J1341.0+3955 & 2.54 $\pm$ 0.18  & 7.42 $\pm$ 2.33 \\
J1419+5423 & 214.9442 & 54.3874 & 0.153 & 3FGL\,J1419.9+5425 & 2.31 $\pm$ 0.06  & 19.6 $\pm$ 2.20 \\
J1427+5409 & 216.8761 & 54.1566 & 0.106 & $-$ & & \\
J1427+3908 & 216.9413 & 39.1423 & 0.165 & $-$ & & \\
J1428+4240 & 217.1361 & 42.6724 & 0.129 & 3FGL\,J1428.5+4240 & 1.58 $\pm$ 0.09  & 2.62 $\pm$ 0.64 \\
J1436+5639 & 219.2408 & 56.6569 & 0.15 & 3FGL\,J1436.8+5639  & 1.99 $\pm$ 0.13  & 4.59 $\pm$ 1.46 \\
J1442+1200 & 220.7012 & 12.0112 & 0.163 & 3FGL\,J1442.8+1200 & 1.80 $\pm$ 0.12  & 3.57 $\pm$ 1.14 \\
J1510+3335 & 227.6716 & 33.5846 & 0.114 & $-$ & &  \\
J1516+2918 & 229.1733 & 29.3026 & 0.13 & $-$ & & \\
J1534+3715 & 233.6967 & 37.2652 & 0.143 & 3FGL\,J1535.0+3721 & 2.11 $\pm$ 0.12  & 5.36 $\pm$ 1.47 \\
J1604+3345 & 241.1938 & 33.7561 & 0.177 & $-$ & & \\
J1647+2909 & 251.8619 & 29.1639 & 0.132 & $-$ & &  \\
\hline
\end{tabular}
\tablefoot{Column 1: the source name; Cols. 2 and 3: Right ascension and declination in J2000; Col. 4: redshift taken from BZCAT; Col. 5: 3FGL name; Col 6: 3FGL photon index; Col 7: 3FGL integrated flux.}
\end{table*}

In \citet[][hereafter \paperi]{Liuzzo2013}, a sample of 42 BL Lacs has been
selected purely on their optical spectra. The sources were selected from the
Roma BZCAT catalogue of known blazars \citep{ref:massaro} to have a measured redshift $z<0.2$ and to be located within the sky area covered by the Sloan Digital Sky Survey \citep[SDSS,][]{ref:sdss}. These two criteria allow us to study with high resolution the bulk of the BL Lac population, including the weakest sources, and grant us the availability of good multi-wavelength coverage. 

The results of a dual-frequency Very Long Baseline Interferometry (VLBI) survey of the sample were presented in \paperi\ of this series. Out of the 42 sources in the sample, 27 sources (64\%) were detected on parsec scales. A radio source can remain undetected in a VLBI observation if it is either below the image sensitivity limit defined by the observation setup or if the bulk emission is distributed over scales substantially larger than the restoring beam of the experiment (typically 1 to 10 parsec in the case of \citealt{Liuzzo2013}).
Among the sources detected by the VLBI observations, some appeared to be of the ``classical'' type, that is,\ bright, compact, with flat spectra and large core dominance (ratio of compact to extended flux). On the other hand, several sources are faint (down to a few mJy and a monochromatic luminosity of $P_\mathrm{r} \sim10^{23.5}$\,W\,Hz$^{-1}$), with moderately steep radio spectra and low core dominance. These properties, in addition to the 15 non-detections, suggest that a significant fraction of BL Lacs are not strongly Doppler boosted.

It is therefore of great importance to study the same population also at high
energy.  Since the start of operation of the Large Area Telescope (LAT) on
board the {\em Fermi Gamma-ray Space Telescope}, the number of BL Lacs
detected in $\gamma$ rays has rapidly increased, reaching 632 in the third LAT
AGN Catalog \citep[3LAC,][]{ref:3LAC}:\ BL Lacs are now the most abundant
population of $\gamma$-ray sources in the sky. Among the sources of our
sample, 20/42 are listed in the 3LAC but it is natural to expect that, in the
course of the ongoing all-sky survey executed by the LAT, more and more of the
sources could be detected in $\gamma$ rays. We list our sources in Table~\ref{tab:table1}.

The main goal of the present paper is therefore to characterise the high-energy
properties of the sample with a larger dataset and an improved LAT-data
analysis. We present a dedicated analysis of 8.5 years of {\em Fermi}\ data for
all the sources in the sample, used to characterise the $\gamma$-ray detection
rate, the $\gamma$-ray flux, and the photon index. The paper is organised as
follows. We present LAT data in Sect.~\ref{sec:data} and the results in
Sect.~\ref{sec:results}. We then discuss the results in
Sect.~\ref{sec:discussion}, and compare them to the 3LAC, the $\gamma$-ray
data above 10 GeV reported in the third catalogue of Hard Fermi-LAT sources
\citep[3FHL;][]{ajello17} and the radio properties of our sample. Finally, we
draw our conclusions in Sect.~\ref{summary}.  

Throughout the paper, we use a $\Lambda$CDM cosmology with $h = 0.68$, $\Omega_m = 0.31$, and $\Omega_\Lambda=0.69$ \citep{Planck16}. The
$\gamma$-ray photon index $\Gamma$ is defined such that $dN/dE \propto
E^{-\Gamma}$. The quoted uncertainties are given at the 1$\sigma$ level, unless otherwise stated.

\begin{table}
\caption{New analysis results for the LAT-detected sources.}
\label{tab:table2}
\centering
\begin{tabular}{cccc}
\hline\hline
Source Name  &  TS & Flux              & Photon Index \\
             &     & (10$^{-9}$ \pflux) & \\
\hline
J0809+3455 & 83    & 1.96 $\pm$ 0.51 & 1.90 $\pm$ 0.09  \\
J0809+5218 & 6513  & 28.26 $\pm$ 1.02 & 1.85 $\pm$ 0.02  \\
J0847+1133 & 242   & 4.11 $\pm$ 0.82 & 1.89 $\pm$ 0.08  \\
J0850+3455 & 109   & 2.63 $\pm$ 0.57 & 1.93 $\pm$ 0.08  \\
J0916+5238 &  34   & 0.54 $\pm$ 0.19 & 1.67 $\pm$ 0.11  \\
J0930+4950 &  28   & 0.35 $\pm$ 0.17 & 1.50 $\pm$ 0.07  \\
J1053+4929 & 216   & 2.71 $\pm$ 0.41 & 1.82 $\pm$ 0.06  \\
J1058+5628 & 5001  & 24.45 $\pm$ 0.91  & 1.92 $\pm$ 0.02  \\
J1120+4212 & 1246  & 4.45 $\pm$ 0.39 & 1.57 $\pm$ 0.03  \\
J1136+6737 &  356  & 2.20 $\pm$ 0.19 & 1.68 $\pm$ 0.03  \\
J1215+0732 &   57  & 1.35 $\pm$ 0.44 & 1.81 $\pm$ 0.11  \\
J1217+3007 & 15910 & 74.12 $\pm$ 1.29  & 1.93 $\pm$ 0.01  \\
J1221+3010 &  3127 & 16.53 $\pm$ 0.98 & 1.70 $\pm$ 0.02  \\
J1221+2813 &  4812 & 43.94 $\pm$ 1.54 & 2.12 $\pm$ 0.02  \\
J1231+6414 &    87 & 1.92 $\pm$ 0.47 & 1.94 $\pm$ 0.09  \\ 
J1253+0326 &   370 & 8.23 $\pm$ 1.23 & 2.02 $\pm$ 0.06  \\
J1341+3959 &    62 & 2.59 $\pm$ 0.56 & 2.16 $\pm$ 0.12  \\
J1419+5423 &  1982 & 31.54 $\pm$ 1.30  & 2.41 $\pm$ 0.03  \\
J1428+4240 &   512 & 2.02 $\pm$ 0.29 & 1.53 $\pm$ 0.05  \\
J1436+5639 &   439 & 4.80 $\pm$ 0.66 & 1.89 $\pm$ 0.06  \\
J1442+1200 &   280 & 3.40 $\pm$ 0.53 & 1.79 $\pm$ 0.06  \\
J1534+3715 &   220 & 7.17 $\pm$ 0.87 & 2.18 $\pm$ 0.06  \\
J1647+2909 &    29 & 0.81 $\pm$ 0.27 & 1.86 $\pm$ 0.11  \\
\hline
\end{tabular}
\tablefoot{Column 1: the source name; Col. 2: TS; Col. 3: flux in the 0.1--300 GeV energy range; Col. 4: $\gamma$-ray photon index.} 
\end{table}

\begin{table}
\caption{Analysis results for the non-LAT-detected sources.}
\label{tab:nodetec}
\centering
\begin{tabular}{cc}
\hline\hline
Source name  &  Upper limit integrated flux \\
             & (10$^{-9}$ \pflux)          \\
\hline
J0751+1730 & 0.23 \\
J0751+2913 & 0.80 \\
J0753+2921 & 0.24 \\
J0754+3910 & 1.34  \\
J0810+4911 & 1.14  \\
J0903+4055 & 1.16  \\
J1012+3932 & 0.38 \\
J1022+5124 & 0.99 \\
J1145$-$0340 & 1.94  \\
J1156+4238 & 0.71 \\
J1201$-$0007 & 2.35  \\
J1201$-$0011 & 2.34  \\ 
J1221+0821 & 2.72  \\
J1257+2412 & 0.51 \\
J1427+5409 & 2.68  \\
J1427+3908 & 0.65 \\
J1510+3335 & 1.08  \\
J1516+2918 & 0.38 \\
J1604+3345 & 2.34  \\
\hline
\end{tabular} 
\tablefoot{Column 1: the source name; Col. 2: 2-$\sigma$ upper limit of the integrated flux in the 0.1--300 GeV energy range.}
\end{table}

\section{{\em Fermi}$-$LAT data}\label{sec:data}

The LAT, the primary instrument onboard the {\em Fermi} $\gamma-$ray observatory, is an electron$-$positron pair conversion telescope sensitive to
$\gamma-$rays of energies from 20~MeV to $>$300~GeV. The LAT consists of a high$-$resolution silicon microstrip tracker, a CsI hodoscopic electromagnetic calorimeter, and an anticoincidence detector for the identification of charged-particle background. The full description of the instrument and its performance can be found in \cite{Atwood2009}. The large field of view ($\sim$2.4~sr) allows the LAT to observe the full sky in survey mode every 3~hours. The LAT point spread function (PSF) strongly depends on both the energy and the conversion point in the tracker, and less so on the incidence angle. For 1~GeV normal$-$incidence conversions in the upper section of the tracker, the PSF 68$\%$ containment radius is 0.8$^{\circ}$  \footnote{https://www.slac.stanford.edu/exp/glast/groups/canda/\\lat\_Performance.htm}.

The {\em Fermi}$-$LAT data presented here have been obtained in a time period of 102 months from 2008 August 5 to 2017 January 31 in the 0.1--300 GeV energy range. During this time, the LAT instrument operated almost entirely in survey mode. The Pass 8 data \citep{atwood13}, based on a complete and improved revision of the entire LAT event-level analysis, were used. The analysis was performed with the \texttt{ScienceTools} software package version v10r0p5. Only events belonging to the `Source' class (\texttt{evclass=128}, \texttt{evtype=3}) were used. We selected only events within a maximum zenith angle of $90^{\circ}$ to reduce contamination from the Earth limb $\gamma$ rays, which are produced by cosmic rays interacting with the upper atmosphere. The spectral analysis was performed with the instrument response functions \texttt{P8R2\_SOURCE\_V6} using a binned maximum-likelihood method implemented in the Science tool \texttt{gtlike}. Isotropic (`iso\_source\_v06.txt') and Galactic diffuse emission (`gll\_iem\_v06.fit') components were used to model the background \citep{acero16}\footnote{http://fermi.gsfc.nasa.gov/ssc/data/access/lat/\\BackgroundModels.html}. The normalization of both components was allowed to vary freely during the spectral fitting.

We analysed a region of interest of $30^{\circ}$ radius centred at the location of each target. We used the coordinates of the $\gamma$-ray sources
associated to the radio source as reported in the third {\em Fermi}-LAT
catalogue of sources \citep[3FGL;][]{acero15}. All BL Lacs of our sample included in the 3FGL have power-law spectra. If the source is not included in the 3FGL we used the radio coordinates of the source. We evaluated the significance of the $\gamma$-ray signal from the source by means of a
maximum-likelihood test statistic (TS) defined as TS = 2$\times$(log$L_1$ - log$L_0$), where $L$ is the likelihood of the data given the model with
($L_1$) or without ($L_0$) a point source at the position of the target \citep[e.g.][]{mattox96}. The source model used in \texttt{gtlike} includes all the point sources from the 3FGL catalogue that fall within $40^{\circ}$ of the target. The spectra of these sources were parametrized by a power-law, a log-parabola, or a super exponential cut-off, according to the model description in the 3FGL catalogue. We also included new candidates within $10^{\circ}$ of our target from the preliminary LAT 8-year point source List (FL8Y) \footnote{https://fermi.gsfc.nasa.gov/ssc/data/access/lat/fl8y/}.

\begin{figure*}
\centering
\includegraphics[width=0.67\columnwidth]{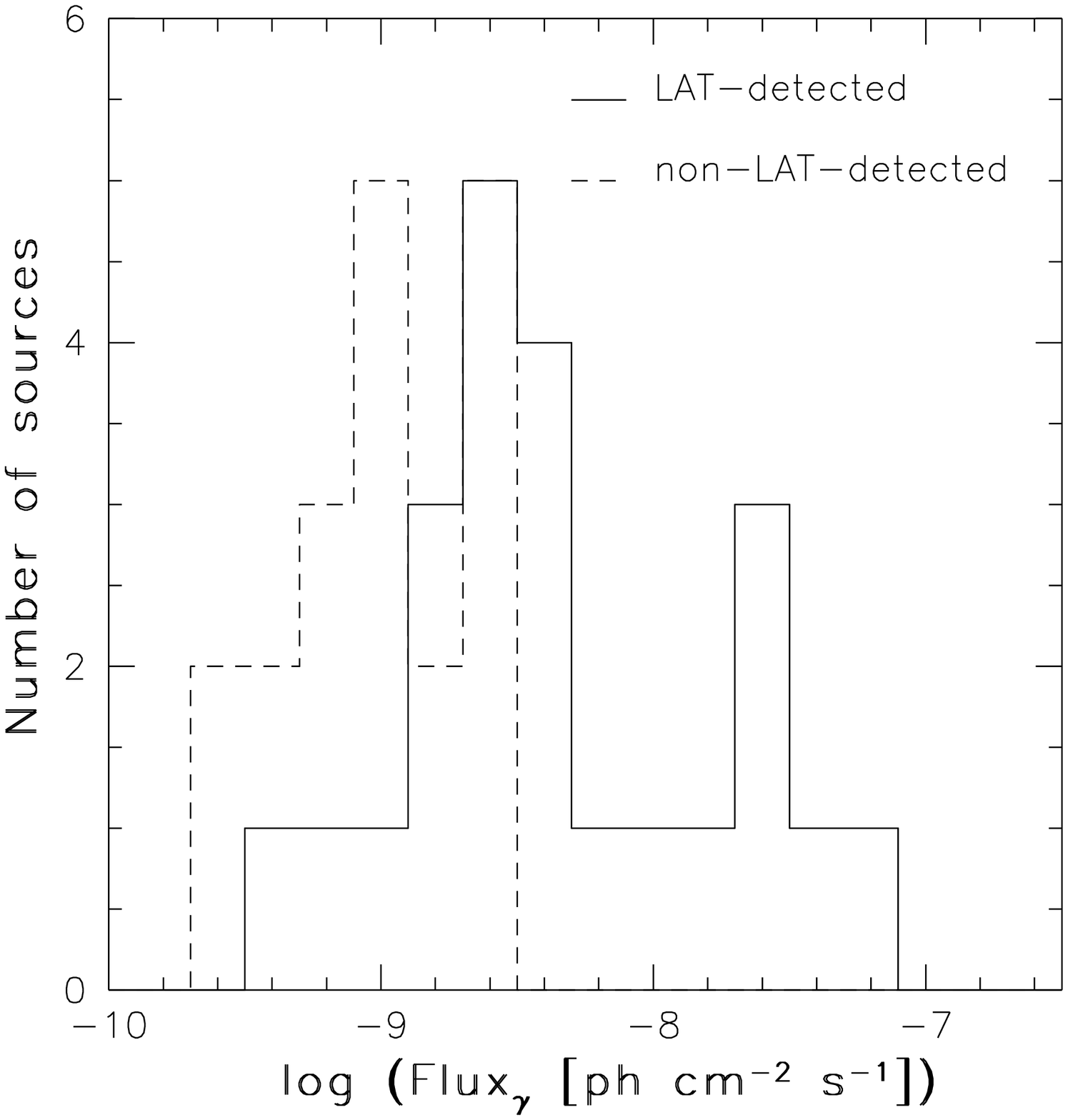} 
\includegraphics[width=0.67\columnwidth]{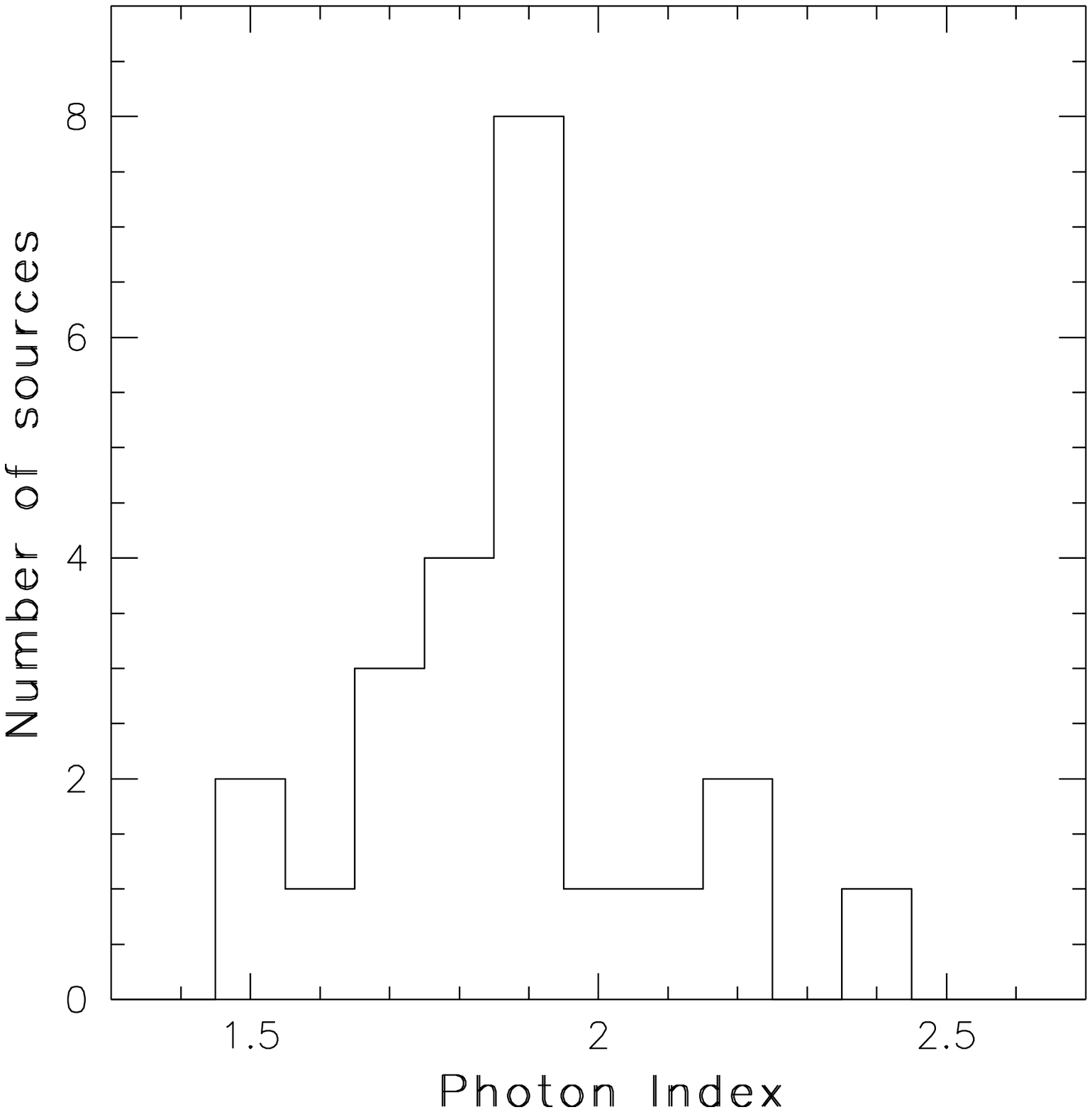} 
\includegraphics[width=0.67\columnwidth]{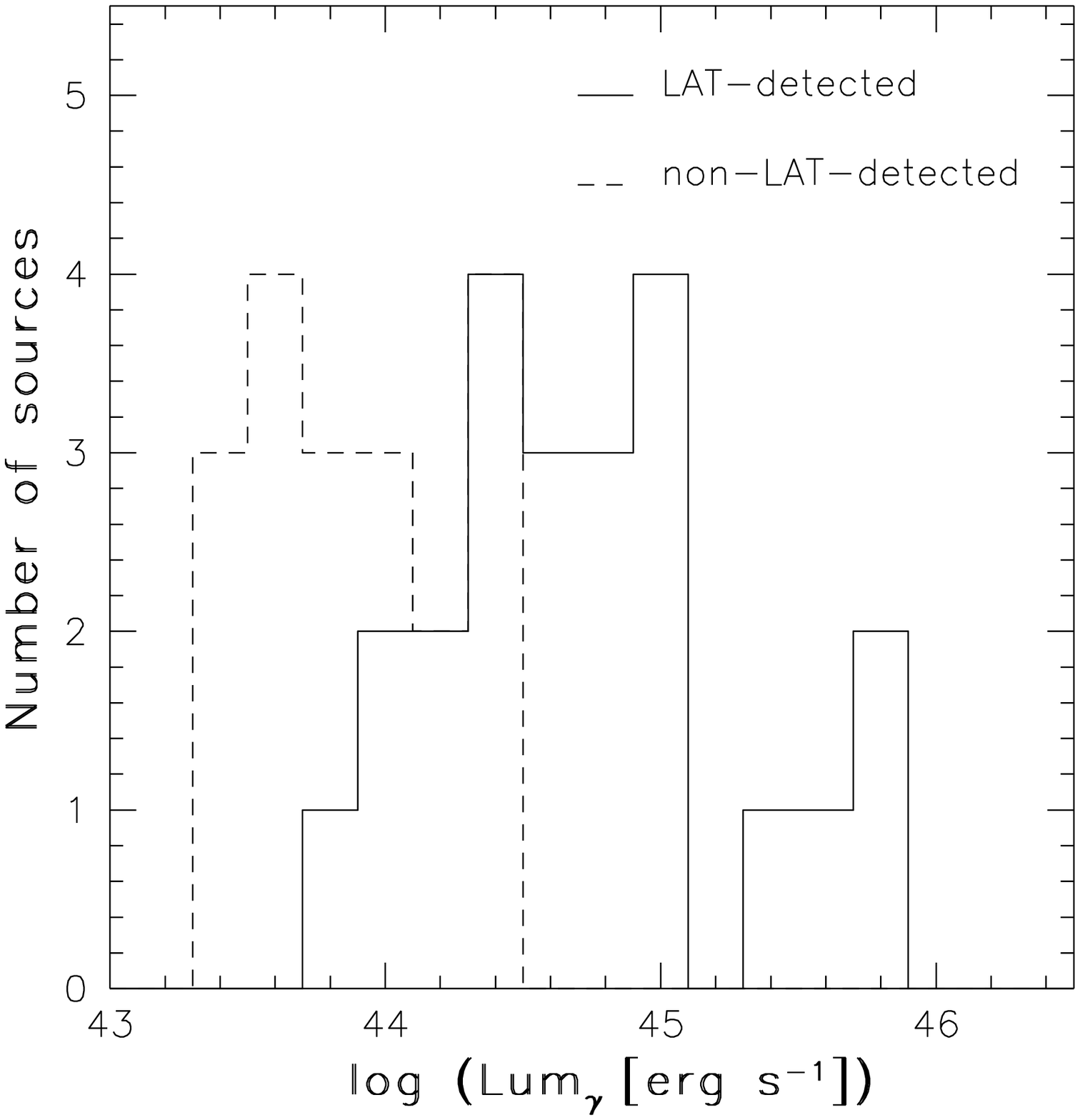} 
\caption{Histogram of the distribution of various quantities obtained by the LAT analysis for the sources in our sample. Short dash line indicates non-LAT-detected sources (TS $<$ 25), for which 2-$\sigma$ upper limits of the flux are considered; red solid line indicates LAT-detected sources (TS $\ge25$). Left panel: integral fluxes in the 0.1--300 GeV energy range; middle: $\gamma$-ray photon index for the 23 LAT-detected sources; right: $\gamma$-ray luminosity in the 0.1--300 GeV energy range.\label{fig:histo}}
\end{figure*}

For the sources under analysis, we modelled the data with a power-law function with both the normalization factor and photon index left free in the
likelihood fit. A first maximum likelihood analysis was performed over the whole period to remove the sources with TS $< 25$ from the model. A second maximum likelihood analysis was performed on the updated source model. In the fitting procedure, the normalization factors and the spectral parameters of the sources lying within 10$^{\circ}$ of the target were left as free parameters. For the sources located between 10$^{\circ}$ and 40$^{\circ}$ from our target, we kept the normalization and the spectral shape parameters fixed to the values from the catalogue. 

\begin{table*}
\caption{Localization of the new $\gamma$-ray sources. }
\label{localization}
\centering
\begin{tabular}{ccccc}
\hline\hline
Source Name  &  RA   & Dec &  95$\%$ error circle radius & Angular separation \\
             &  deg  & deg &  deg                        & deg                \\
\hline
J0916+5238 & 139.189 & 52.647 & 0.038 & 0.018 \\
J1215+0732 & 183.773 &  7.519 & 0.039 & 0.027 \\
J1647+2909 & 251.861 & 29.170 & 0.042 & 0.006 \\
\hline
\end{tabular} 
\tablefoot{Column 1: the source name; columns 2 and 3: Right ascension and declination of the $\gamma$-ray source; column 4: LAT 95$\%$ error circle radius; column 5: angular
 separation between the coordinates of the $\gamma$-ray source and those of the radio counterpart.} 
\end{table*}

For the new detected sources, we ran a point source localization using the \texttt{gtfindsrc} tool over the photons with $E$ $>$ 1 GeV extracted during the whole period. We also tried to parameterize the spectrum of these sources with a log parabola model, but the fit with a log parabola did not provide a statistically significant improvement with respect to a power-law. These three sources are included in the FL8Y list.   

\section{Results of the {\em Fermi}-LAT data analysis} \label{sec:results}

\begin{figure}
\centering
\includegraphics[width=0.9\columnwidth]{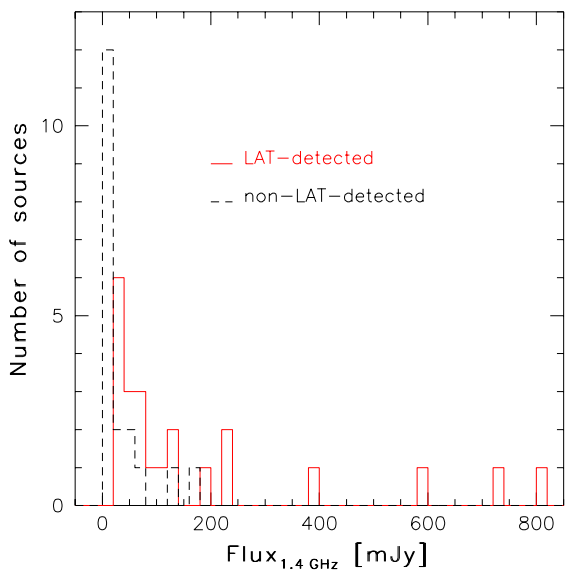}
\caption{Histogram of the NVSS flux at 1.4 GHz of the sources in our sample. Short dash line indicates non-LAT-detected sources (TS $<$ 25), red solid line indicates LAT-detected sources (TS $\ge25$).}\label{NVSS}
\end{figure}

Here, the $\gamma$-ray detection rate, fluxes, and photon indexes of the sources in our sample are reported. We detected 23 out of 42 sources with TS $\ge25$ (corresponding to a $\sim$ 4.6-$\sigma$ confidence level for two degrees of freedom). The results of our analysis for the LAT-detected sources are summarised in Table~\ref{tab:table2}. For the sources not detected in our $\gamma$-ray analysis, a 2-$\sigma$ upper limit of the integrated flux (assuming a photon index $\Gamma$ = 2) is reported in Table~\ref{tab:nodetec}. For two sources the likelihood analysis results in a TS between 10 and 25: J0754$+$3910 (TS = 12) and J0810$+$4911 (TS = 11). Adding more data, these sources could be detected by {\em Fermi}-LAT in the future.

\begin{figure*}
\centering
\includegraphics[width=\columnwidth]{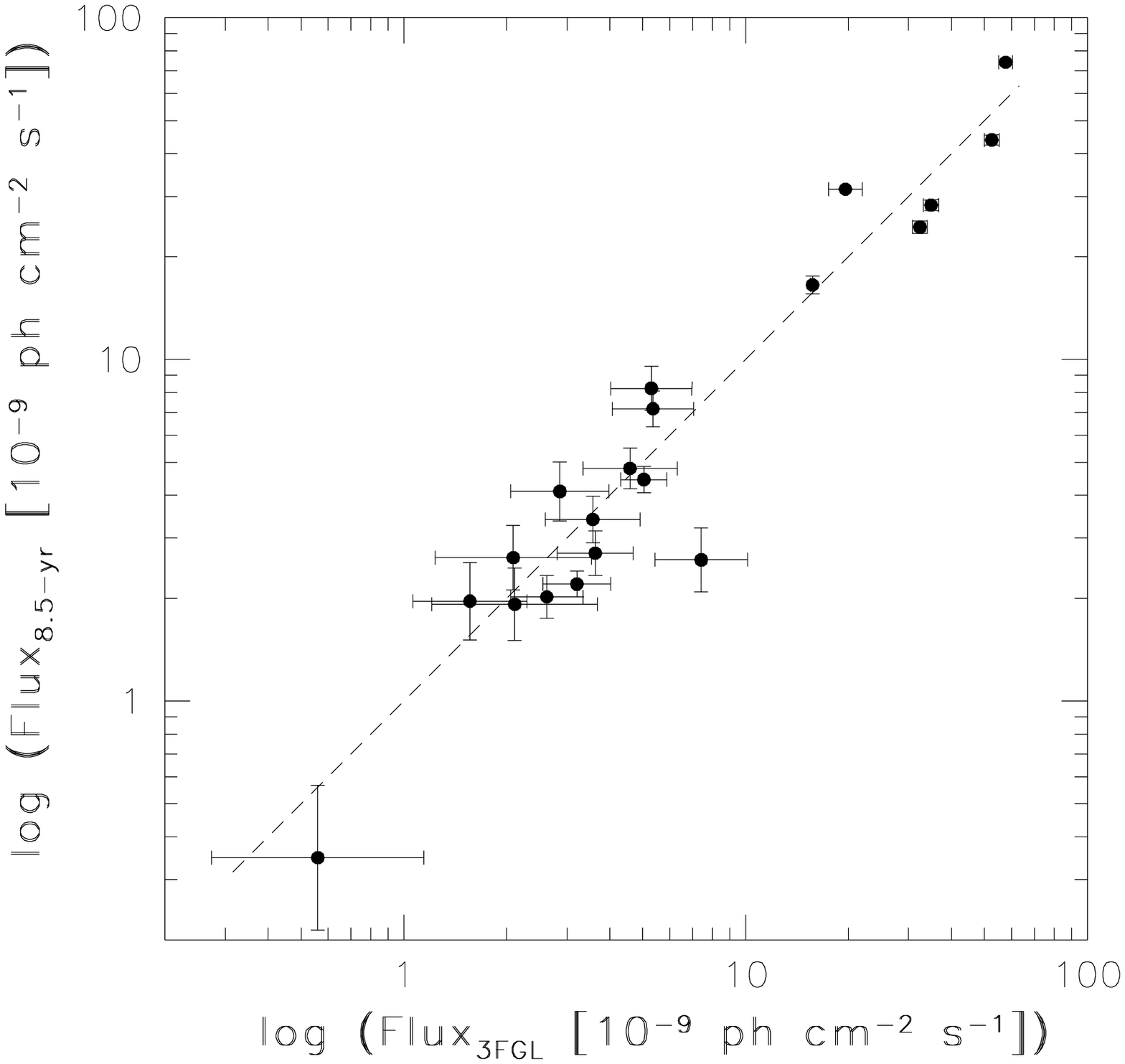} 
\includegraphics[width=\columnwidth]{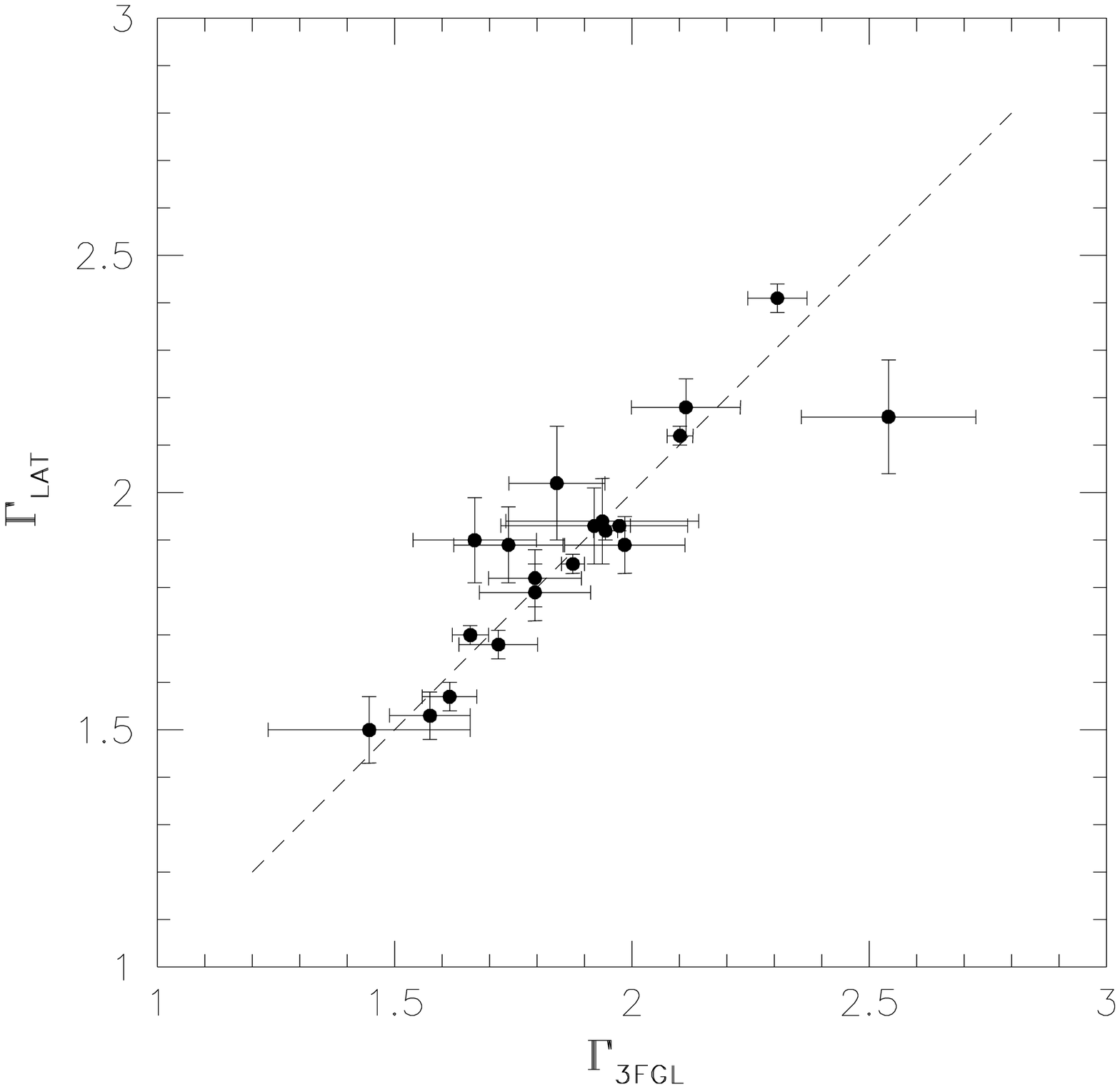} 
\caption{Plot of the $\gamma$-ray flux (left panel) and photon index (right panel) obtained in this work versus the values reported in the 3FGL. A one-to-one reference short dashed line is plotted.} \label{fig:phvsflulum}
\end{figure*}

\begin{figure}
\centering 
\includegraphics[width=\columnwidth]{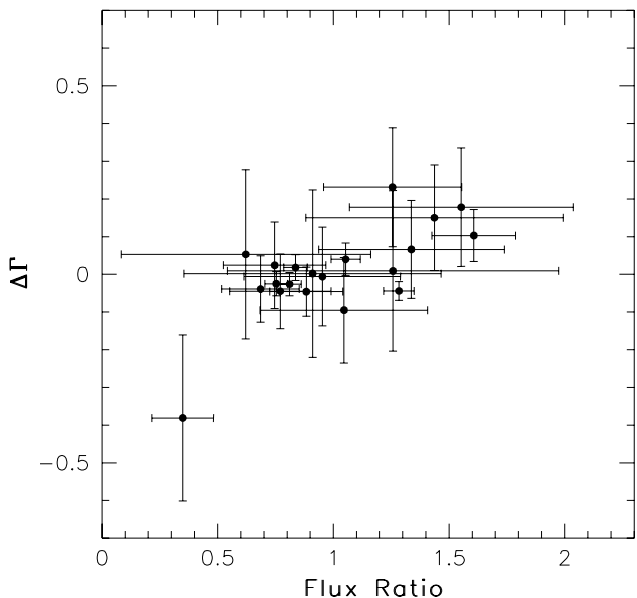}
\caption{Photon index difference between 8.5-yr and 3FGL values versus flux ratio (Flux$_{\rm\,8.5-yr}$/Flux$_{\rm\,3FGL}$) is shown.}\label{fig:phvsflulum2}
\end{figure}

We confirm the detection of all the sources previously reported in the
3FGL. In addition, three $\gamma$-ray sources not in the 3FGL and associated with
BL Lacs in our sample have been revealed, raising the overall detection rate
for the sample from 48\% to 55\%. We ran a localization of the three new detected
sources, and the results are reported in Table \ref{localization}. This
further analysis confirms a strict spatial association between the
$\gamma$-ray source and the radio counterpart associated with the BL Lac. Two
out of three sources (J0916+5238 and J1215+0732) are included in the Brazil
ICRANet Gamma-ray Blazar (1BIGB) catalogue \citep{arsioli17}.

In Fig.~\ref{fig:histo}, we show the distribution of the integral fluxes in the 0.1--300 GeV energy range (left panel), the photon indexes (middle panel), and the $\gamma$-ray luminosity (right panel) for the sources in our sample including 2-$\sigma$ upper limits of the flux and corresponding luminosity for non-LAT-detected sources. The integral fluxes of the LAT-detected sources span the range between $3.5\times 10^{-10}$ and $7.4\times10^{-8}$ \pflux, with a peak at $\sim$ $4\times$ 10$^{-9}$ \pflux. The photon index distribution has a peak at around $\Gamma=1.7-2.0$, with only four sources having values in the range 2.1--2.4. The average photon index of the LAT-detected sources in our sample is $\langle \Gamma \rangle = 1.87 \pm 0.06$. The three new detected sources have fluxes below $1.4\times10^{-8}$ \pflux, luminosity below 2.5$\times$10$^{44}$ erg s$^{-1}$, and photon index 1.7--1.9.

For the sources not detected by LAT the 2-$\sigma$ upper limit of the flux ranges between $0.2\times 10^{-9}$ and $2.7\times 10^{-9}$ \pflux. The lack of detection of these sources in $\gamma$ rays is in agreement with the LAT sensitivity limit\footnote{http://www.slac.stanford.edu/exp/glast/groups/canda/\\lat\_Performance.htm} estimated for 10 years of observations assuming a photon index $\Gamma$ = 2, that is, $1-2\times10^{-9} \pflux$. 

All sources with luminosity $>$ 5 $\times$10$^{44}$ erg s$^{-1}$ have been detected in $\gamma$ rays by LAT. On the other hand, no sources have been detected in $\gamma$ rays below 8$\times$10$^{43}$ erg s$^{-1}$ so far, suggesting that a significant portion of the bulk of the BL Lac object population has not been probed yet with {\em Fermi}-LAT. There is no clear distinction between the distribution of the LAT-detected and
non-LAT-detected sources in the range of luminosity 1--5 $\times$10$^{44}$ erg s$^{-1}$.

The LAT-detected and non-LAT-detected sources are equally distributed in redshift (0.066 $< z <$ 0.199 for the former,  0.096 $< z <$ 0.187 for the
latter), suggesting that the redshift has no impact on the $\gamma$-ray detection in our sample. A Kolmogorov-Smirnov (K-S) test \citep{press92} does not show a significant difference between the redshift distributions of LAT-detected and non-LAT-detected BL Lacs in our sample. 

If we consider the distribution of flux density at 1.4 GHz obtained by NRAO VLA Sky Survey (NVSS), it is evident that the radio flux densities of LAT-detected sources is usually higher than those of the non-LAT detected sources in our sample (Fig.~\ref{NVSS}). However there are 12 LAT-detected BL Lacs with flux density $<$ 85 mJy \citep[i.e. the flux limit of the VLBA Imaging and Polarimetry Survey, VIPS;][]{helmboldt07}, showing that our sample is representative of a population of LAT-detected BL Lacs for which pc-scale radio properties were unexplored in detail up to our VLBA observations.

\section{Discussion} \label{sec:discussion}

The project described in this work is aimed to improve our knowledge of the BL Lac population, through VLBI and {\em Fermi}-LAT observations of a sample of BL Lacs, independently of their $\gamma$-ray properties and with multi-wavelength information available on all targets.

\subsection{Comparison with 3LAC}

For the 20 sources already reported in the 3LAC, we show in Fig.~\ref{fig:phvsflulum} the plot of the flux and photon index obtained in this work with 8.5 years of Pass 8 LAT data with respect to the values reported in the 3FGL for 4 years of Pass 7 reprocessed LAT data. In Fig.~\ref{fig:phvsflulum2}, we show the ratio of the $\gamma$-ray fluxes obtained in 8.5 years and 4 years and the photon index variation in the two periods of LAT observations. The fluxes are in general in agreement, with a mean value of the ratio between the value over 8.5 years and that reported in the 3FGL of 1.01 $\pm$ 0.28. The most prominent outliers are: J1419+5423, whose flux has increased by 60\%, and J1341+3959, whose flux decreased by a factor of three when considering 8.5 years of LAT data. This is in agreement with the moderate variability usually observed in the LAT energy range for BL Lacs \citep[e.g.][]{ref:3LAC}.

In the right panel of Fig.~\ref{fig:phvsflulum}, we plot the 8.5-yr photon
index versus the values reported in the 3FGL. The photon indexes are generally
consistent, having $\langle \Delta \Gamma \rangle = \langle
\Gamma_\mathrm{8.5-yr}-\Gamma_\mathrm{3FGL} \rangle$ = 0.08 $\pm$ 0.16. Only
for J1341+3959 there is a hint of change of the photon index, from $\Gamma_\mathrm{3FGL}$ = 2.54 $\pm$ 0.18 to $\Gamma_\mathrm{8.5-yr}$ = 2.16 $\pm$ 0.12.

In general, variations in $\gamma$-ray flux and in photon index are not correlated. Neither softer-when-brighter nor harder-when-brighter trends are found, as shown by the comparison between the $\gamma$-ray flux ratio and the photon index variation presented in Fig.~\ref{fig:phvsflulum2}, with the exception of J1341+3959 that shows a softer-when-brighter trend. 

\begin{figure}
\centering
\includegraphics[width=\columnwidth]{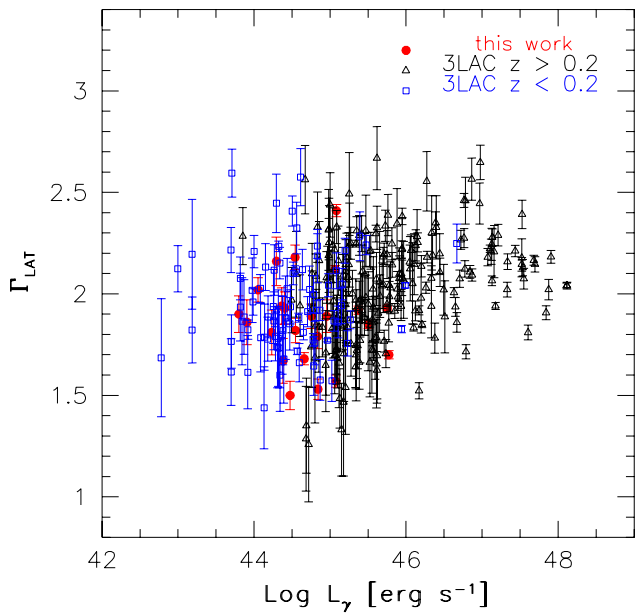}
\caption{Photon index vs. $\gamma$-ray luminosity in the 0.1--300 GeV energy range. Red stars: BL Lacs detected from our sample, black open triangles: BL Lacs in the 3LAC with $z > 0.2$, blue open squares: BL Lacs in the 3LAC with $z < 0.2$.}\label{fig:Lum_Gamma}
\end{figure}

In Fig.~\ref{fig:Lum_Gamma}, we compare the $\gamma$-ray photon index and luminosity obtained over 8.5 years for the 23 LAT-detected BL Lacs in our
sample with values in the 3LAC for the other BL Lacs with known redshift, separating sources with $z < 0.2$ and $z > 0.2$. The average apparent
$\gamma$-ray isotropic luminosity in the 0.1--300 GeV energy range of the LAT-detected sources in our sample ranges between 6 $\times$ 10$^{43}$ and 6 $\times$ 10$^{45}$ erg s$^{-1}$, which are typical luminosities for BL Lacs and within the range of values observed for the other BL Lacs with $z < 0.2$ in the 3LAC \citep[6$\times$10$^{42}$--5$\times$10$^{46}$ erg s$^{-1}$;][]{ref:3LAC}. In the same way, the average photon index of the
LAT-detected sources ranges between 1.5 and 2.4, in agreement with what is observed for BL Lacs in the 3LAC. In particular, the average photon index of our sample, $\langle \Gamma_{\rm\,8.5-yr} \rangle$ = 1.87 $\pm$ 0.06, is compatible with the average photon index reported in the 3LAC for the other BL Lacs with $z < 0.2$, $\langle \Gamma_{\rm\,3LAC\,, z < 0.2} \rangle$ = 1.95 $\pm$ 0.11. This suggests that the $\gamma$-ray spectra of the LAT-detected BL Lacs in our sample are similar to the spectra of the other BL Lacs in the 3LAC, in particular at $z <0.2$. Further observations at other wavelengths are needed to investigate differences and similarities between the properties of this sample and those of the other BL Lacs detected by LAT.

Based on the frequency of the synchrotron peak ($\nu_{peak}^{syn}$), BL Lacs can be divided into low-synchrotron-peaked (LSP; $\nu_{peak}^{syn} < 10^{14}$ Hz), intermediate-synchrotron-peaked (ISP; $10^{14} < \nu_{peak}^{syn} < 10^{15}$ Hz), and high-synchrotron-peaked (HSP; $\nu_{peak}^{syn} > 10^{15}$ Hz) BL Lacs \citep{Abdo2010}. Based on the $\nu_{peak}^{syn}$ values reported in the 3LAC, among the LAT-detected sources in our sample 16 are HSP, 3 ISP, and 1 LSP BL Lac. Two out of the three new LAT-detected sources are classified as HSP BL Lac in the second {\em Wide-field Infrared Survey Explorer (WISE)} HSP catalogue \citep[2WHSP;][]{chang17}. Among the 19 BL Lacs not detected by LAT, 11 are classified as HSP or HSP candidates in
the 2WHSP; therefore in our sample 30 of 42 sources (71\%) are classified as HSP or candidate HSP BL Lacs. Low-synchrotron-peaked sources are usually the brightest BL Lacs
detected by LAT. The rising part and high-energy peak of the SED of HSP BL Lacs lies within the LAT energy range, favouring their detection by
LAT, as shown by the large number of HSP BL Lacs in the 3LAC. Among the 30 HSP BL Lacs in our sample, 19 sources (63\%) are detected in $\gamma$ rays, while only 4 out of the 12 (33\%) ISP/LSP BL Lacs have been detected by {\em Fermi}-LAT so far. 

\begin{figure}
\centering
\includegraphics[width=\columnwidth]{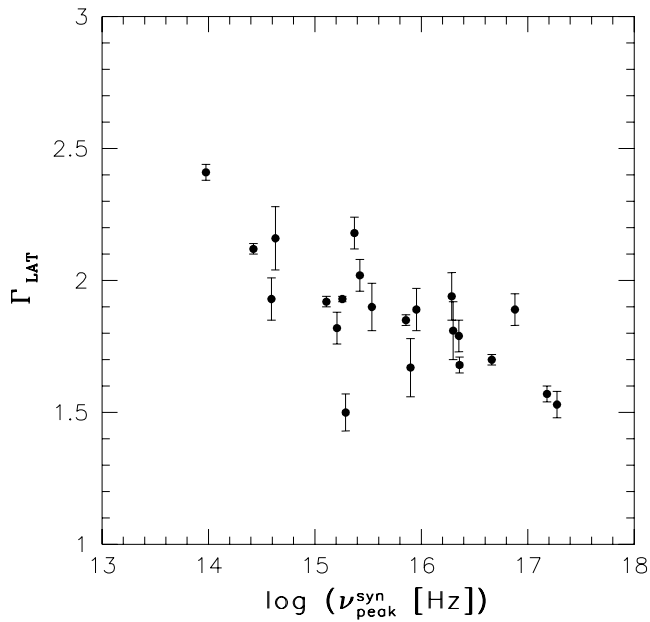}
\caption{$\gamma$-ray photon index vs. synchrotron peak frequency of the 22 LAT-detected sources in our sample with $\nu_{peak}$ estimated.}\label{sequence} 
\end{figure}

An anti-correlation between the synchrotron peak frequency and the LAT photon index measured over 8.5 years is observed for the LAT-detected BL Lacs in our sample (Fig.~\ref{sequence}), as already observed for the whole sample of blazars reported in the 3LAC \citep{ref:3LAC}, and expected by the blazar sequence proposed by \citet{Fossati1998} and \citet{Ghisellini1998}.

Only three HSP sources of our sample (SWIFT J1136$+$6738, SWIFT J1221$+$3012, and SWIFT J1428$+$4234) with very high synchrotron peak frequency
(i.e. 10$^{17}$--10$^{18}$ Hz) are included in the 105-month {\em Swift}-BAT
catalogue \citep{oh18}. These three sources have an X-ray photon index in the
14--195 keV energy range of 2.3--2.9 and a corresponding $\gamma$-ray photon
index of 1.5--1.7, indicating that we are observing the declining part of the
synchrotron emission in hard X-rays and the rising part of the inverse Compton emission in $\gamma$ rays. 

\subsection{Comparison with 3FHL}

\begin{figure}
\centering
\includegraphics[width=\columnwidth]{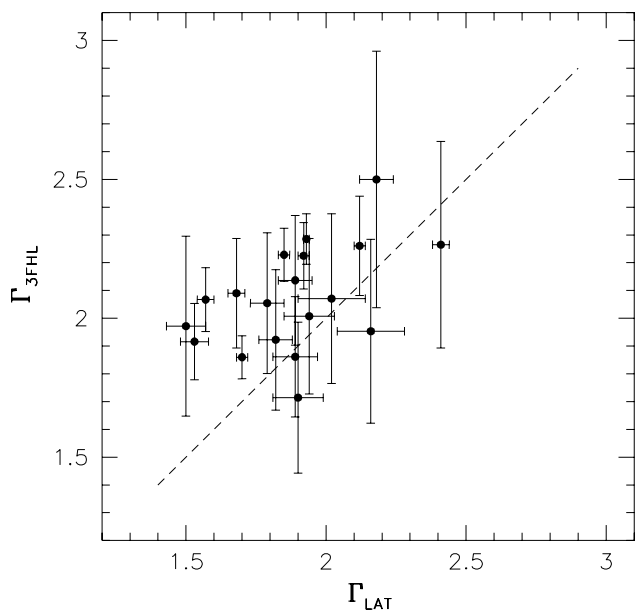}
\caption{3FHL photon index vs. photon index estimated over 8.5 years in the 0.1--300 GeV energy range. A one-to-one reference short dashed line is plotted.}\label{3FHL_ph_comparison}
\end{figure}

In addition to the LAT catalogues with low-energy threshold of 0.1 GeV, three
hard-source catalogues have been released, the most recent one being the 3FHL,
based on 7 years of data analysed in the 10 GeV--2 TeV energy range. The 3FHL
contains 1556 objects and takes advantage of the improvement provided by Pass
8 data by using the PSF-type event classification. The vast majority of
detected sources (79\%) are associated with extragalactic counterparts at
lower energies; in particular, BL Lacs are the dominant extragalactic
population (i.e. 750 objects). Among the 23 BL Lacs detected by {\em
  Fermi}-LAT in the 0.1--300 GeV energy range, 19 are included in the
3FHL. This is not surprising, considering that most of the LAT-detected
sources in our sample are HSP, with the high-energy peak within the energy
range covered by the 3FHL. All sources of our sample included in the 3FHL are
also detected in the 0.1--300 GeV energy range. Among the 19 sources in the
3FHL, 9 are also detected at very high energy (VHE; E $>$ 100 GeV) by the
current Imaging Atmospheric Cherenkov Telescopes \footnote{TeVCat v3.400:
  http://tevcat.uchicago.edu/}. Moreover, 11 out 19 3FHL sources are included
also in the 2FHL \citep[][]{ackermann16}, where a higher energy threshold of
50 GeV was considered for the analysis. Four 2FHL sources in our sample (2FHL J0809.5+3455, 2FHL J1053.5+4930, 2FHL J1058.5+5625, 2FHL J1120.8+4212) are not detected at VHE and could be good targets for the current Cherenkov Telescopes.

The difference between the 3FHL photon index and the photon index estimated in this work in the 0.1--300 GeV energy range ranges between $-$0.21 and 0.50. In Fig.~\ref{3FHL_ph_comparison} we compare the 3FHL photon index with the LAT photon index in the 0.1--300 GeV energy range. The average 3FHL photon index of the 19 sources in our sample is $\langle \Gamma_{\rm\,3FHL} \rangle = 2.07 \pm 0.23$. As a comparison the average photon index estimated in this work in the 0.1--300 GeV energy range is $\langle \Gamma_{\rm\,8.5-yr} \rangle = 1.88 \pm 0.05$, indicating a small difference between the photon index estimated in the two energy ranges. The difference between the two photon indices of the 19 LAT-detected sources in both samples is -0.21 $<$ $\Delta\Gamma_{\rm\,3FHL} \equiv (\Gamma_{\rm\,3FHL}-\Gamma_{\rm\,8.5yr})$ $<$ 0.50, suggesting that the average $\gamma$-ray spectrum of these sources below 300 GeV connects rather smoothly with the spectrum in the 10 GeV--2 TeV energy range. For five sources, even considering the uncertainties, the 3FHL photon index is larger than the 3FGL photon index, suggesting a possible curvature in the overall $\gamma$-ray spectrum. Removing these five sources, the average photon index estimated in the 0.1--300 GeV energy range ($\langle \Gamma_{\rm\,8.5-yr} \rangle = 1.99 \pm 0.06$) and in the 10 GeV--2 TeV energy range ($\langle \Gamma_{\rm\,3FHL} \rangle = 2.11 \pm 0.25$) are still compatible. 
In case of sources with $\Delta\Gamma_{\rm\,3FHL} < 0$, and therefore a harder 3FHL photon index value with respect to the 0.1--300 GeV value, the two photon indexes are compatible within the uncertainties, as shown in Fig.~\ref{3FHL_ph_comparison}. This is an indication of no significant break or softening in the observed spectra between 0.1 GeV and 2 TeV. This should be related also to the small effect of absorption due to interactions with the extragalactic background light \citep[EBL,][]{dominguez11} on the spectra of BL Lacs at low redshift, such as those in our sample \citep[see e.g.][]{dominguez15}.

\begin{figure}
\centering
\includegraphics[width=\columnwidth]{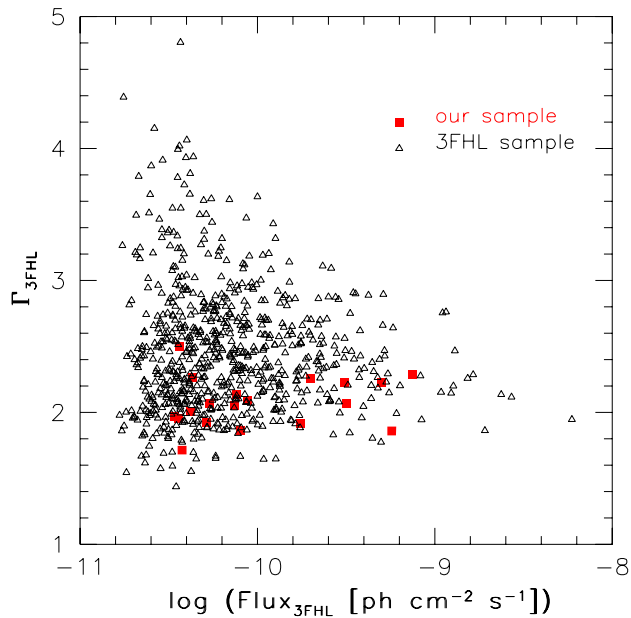}
\caption{Photon index vs. flux of the BL Lacs included in the 3FHL. Red squares show the BL Lacs included in our sample.}\label{3FHL}
\end{figure}

In Fig. \ref{3FHL}, we plot the $\gamma$-ray photon index versus flux in the 10 GeV -- 2 TeV energy range of the BL Lacs included in the 3FHL, with the 19 BL Lacs of our sample shown with filled red squares. The sources in our sample detected above 10 GeV have a photon index $\Gamma_{\rm\,3FHL} <$ 2.6, independently from the flux. Therefore, the photon index seems to be more important than the flux for detecting sources at high energies, at least when considering a sample in the same bin of redshift, that is, with similar effect of absorption due to interactions with the EBL. All the 3FHL sources in our sample are good candidates for future observations with the Cherenkov Telescope Array \citep{acharya18}.

\subsection{Comparison with VLBI data at 8 and 15 GHz}

Radio and $\gamma$-ray emission are strongly correlated in blazars \citep[e.g.][]{ghirlanda10,Ackermann2011,richards14,mufakharov15,bock16}. However, given the small field of view of VLBI arrays, large surveys with milliarcsecond angular resolution are not available and for parsec scale studies one has to resort to dedicated pointed observations. \citet{Linford2012} present VLBI data for a large sample of $\gamma$-ray sources, including as many as 95 BL Lacs; interestingly, they find that LAT-detected BL Lacs do not differ significantly from a control sample of non-LAT-detected BL Lacs in terms of brightness and morphology. 

\begin{figure}
\centering
\includegraphics[width=\columnwidth]{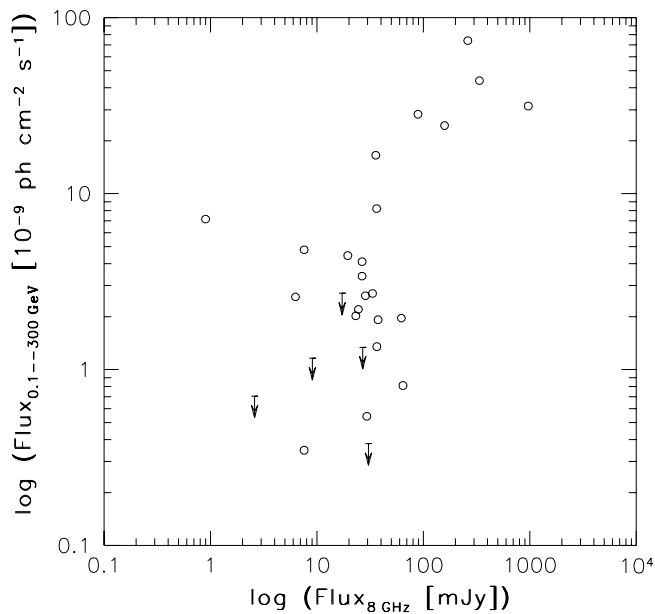}
\caption{Flux density at 8 GHz vs. $\gamma$-ray flux estimated in the 0.1--300 GeV energy range over 8.5 years of LAT data. Arrows show 2-$\sigma$ upper limits of the $\gamma$-ray flux.}\label{Gamma_radio}
\end{figure}

The results of dual frequency (8 and 15 GHz) VLBA observations for the sources in our sample were presented in \citetalias{Liuzzo2013}. Compact components were found for 27 sources. The presence of $\gamma$-ray emission
and of a compact radio component seem to be well connected: 21 of the 23
LAT-detected sources (91\%) reported in this paper have a VLBI component with
respect to 6 of the 19 non-LAT-detected sources (32\%). The only LAT-detected
sources undetected at parsec scale are J0847$+$1133 and J1534$+$3715. However,
the lack of detection of J0847$+$1133 is likely due to observational problems
during VLBI observations reported in \citetalias{Liuzzo2013}. Nevertheless, a
compact emission has been observed with the European VLBI Network (EVN) at 5
GHz \citep{mantovani15} and with VLBA at 8 GHz \citep{bourda10} for this
source. On the other hand, no radio emission has been revealed at both 8 GHz
and 15 GHz for J1534$+$3715 with VLBA. It is worth noticing that this source
has the lowest NVSS flux densities at 1.4 GHz among the LAT-detected sources
in our sample, suggesting that the radio emission is not concentrated in the
core region but extended emission is present on scales which are not probed by VLBI arrays.

The six non-LAT-detected sources with a compact component detected by VLBA have
a 8 GHz flux density in the range 2.6--30.0 mJy with respect to the range of
flux densities of 6.0--969.5 mJy for the LAT-detected sources. This result suggests that these sources are in the low-values part of the distribution of 8 GHz flux
density in our sample. Comparing the $\gamma$-ray flux in the 0.1--300 GeV
energy range with the 8 GHz flux density of the 27 sources for which parsec
scale emission is detected by VLBA (Fig. \ref{Gamma_radio}), a connection
between the radio and $\gamma$-ray emission is evident, in agreement with the
radio/$\gamma$-ray flux correlation reported in \citet{Ackermann2011}. We calculate a Pearson correlation coefficient between 8 GHz flux densities and $\gamma$-ray fluxes and obtain $R$ =  0.5746, indicating a moderate positive correlation ($P$-value = 0.001384).

\begin{figure}
\centering
\includegraphics[width=\columnwidth]{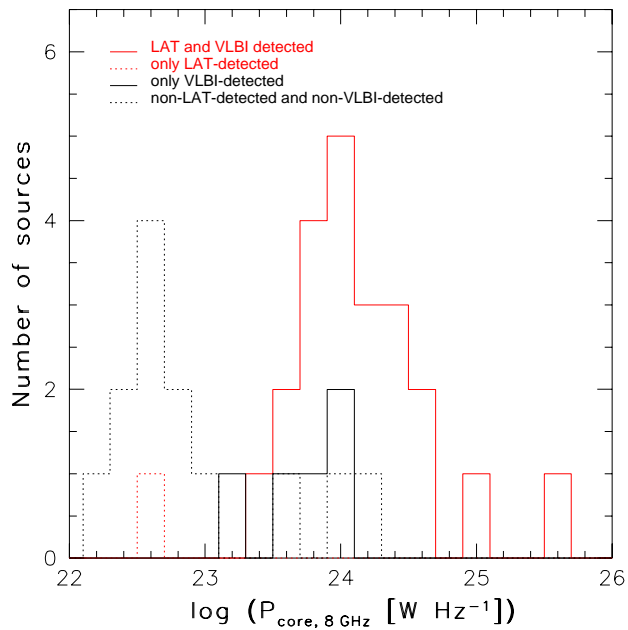}
\caption{Histogram of the core power at 8 GHz of the sources in our sample. Solid line indicates that the source is detected and short-dash line shows the upper limit. Red lines refer to LAT-detection and black lines refer to VLBI-detection.}\label{power8GHz}
\end{figure}

\begin{figure}
\centering
\includegraphics[width=\columnwidth]{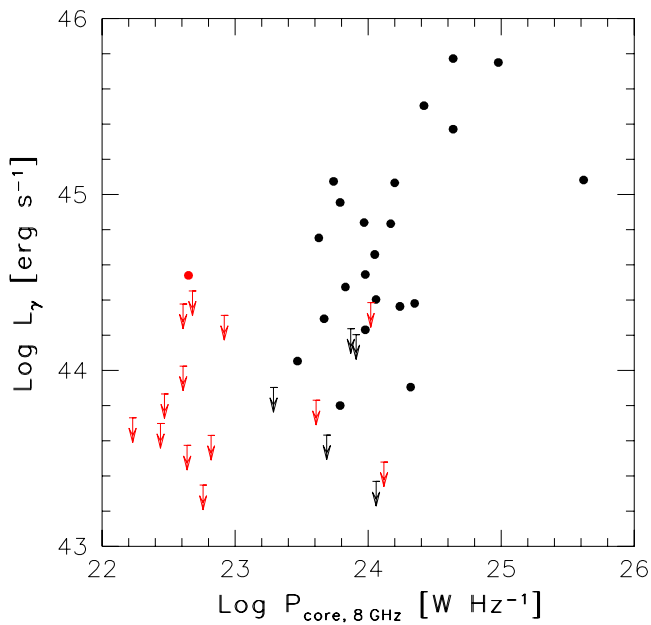}
\caption{Luminosity of the core at 8 GHz versus $\gamma$-ray luminosity in the
  0.1--300 GeV energy range for the sources in our sample. Filled circles indicate LAT-detected sources, while arrows show upper limits of the
  $\gamma$-ray luminosity. If only an upper limit on the radio luminosity has
  been estimated, the data are reported in red.}\label{LumGLumR}
\end{figure} 

In Fig.~\ref{power8GHz} we compare the VLBA core power at 8 GHz for 41 sources
(J0751$+$2913 has not been observed at 8 GHz). Upper limits are reported for
14 sources \citepalias[see][Table 5 for details]{Liuzzo2013}; all of them are
non-LAT-detected sources, except for J1534$+$371. The LAT-detected sources
have on average a higher core power at 8 GHz, P$_{\rm\,core,\,8\,GHz}$, $\langle P_{\rm\,core,\rm\,8\,GHz}^{\rm\,LAT} \rangle$ = 1.3$\times$10$^{24}$ W
Hz$^{-1}$, with respect to non-LAT-detected sources, $\langle P_{\rm\,core,\,8\,GHz}^{\rm\,non-LAT} \rangle$ = 1.4$\times$10$^{23}$ W
Hz$^{-1}$. All sources with $P_{\rm\,core,\rm\,8\,GHz}$ $>$ 2 $\times$10$^{24}$ W Hz$^{-1}$ have been detected in $\gamma$ rays by LAT. On
the other hand, only one source with $P_{\rm\,core,\rm\,8\,GHz}$ $<$ 3 $\times$10$^{23}$ W Hz$^{-1}$, J1534$+$3715, has been detected by
LAT. This indicates that LAT-detected sources are the most luminous radio sources of our sample on parsec scale. Moreover, it suggests that there is a
population of BL Lacs that lack a VLBI core, have a low $\gamma$-ray emission, and likely have a small Doppler factor. This is even more
evident by comparing the $\gamma$-ray luminosity in the 0.1--300 GeV energy range with the radio luminosity of the core at 8 GHz (Fig.~\ref{LumGLumR}): the bottom-left part of the plot is populated by sources with a low luminosity in both the radio and $\gamma$-ray bands. All the sources that lie in that part of the luminosity-luminosity plot are not yet detected by {\em Fermi}-LAT, except for J1534$+$3715.

\begin{figure}
\centering
\includegraphics[width=\columnwidth]{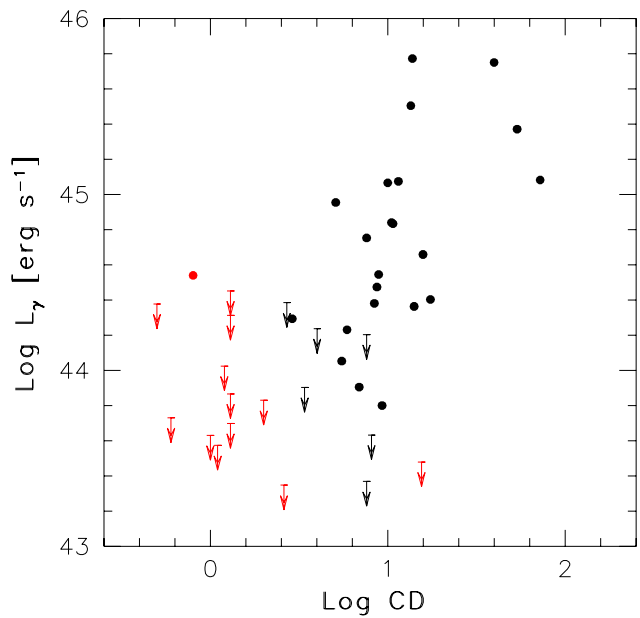}
\caption{Core dominance vs. $\gamma$-ray luminosity in the 0.1--300 GeV energy range for the sources in our sample. Filled circles indicate LAT-detected sources, while arrows show upper limits of the $\gamma$-ray luminosity. If only an upper limit on the CD has been estimated, the data are reported in red.}\label{CD}
\end{figure} 

On the basis of the radio properties (i.e. source compactness, Core Dominance,
CD\footnote{CD is defined as ratio between the observed core radio power and the estimated unbeamed total radio power at low frequency}, source morphology and radio spectrum) in \citetalias{Liuzzo2013},
all the sources have been classified in three groups: Doppler-dominated (DD, 23 cases), lobe-dominated (LD, 11 cases), and undetermined type (U, 8 cases) sources. In DD sources, the luminosity of the core is much larger than the luminosity of the extended component, implying relativistic jets and small viewing angles; LD sources have less prominent or undetected cores, and significant extended emission; U sources are undetected with VLBI but their overall flux density is so low that it does not provide stringent limits on the ratio between nuclear
and extended emission. Among the 23 LAT-detected sources, 20 are classified as
DD, and 3 as LD (J0847$+$1133, J1341$+$3959, and J1534$+$3715). All three newly
detected $\gamma$-ray sources are classified as DD. European VLBI Network and VLBA observations
at 5 GHz and 8 GHz, respectively, and a CD of 5.1 suggests an intermediate
classification of J0847$+$1133 between LD and DD. Faint radio emission has
been detected with VLBA observations of the other two LAT-detected sources
classified as LD. On the other hand, three sources classified as DD for their
radio properties (J0754$+$3910, J0903$+$4055, J1516$+$2918) are not detected
in $\gamma$ rays by LAT up to now. These three sources have a relatively low CD
(7.6--8.0) with respect to the LAT-detected sources classified as DD
(5.5--72.4). Considering that CD is a strong indication of Doppler beaming of
the emission, this suggests that the DD sources not detected by LAT have a lower
Doppler factor with respect to the LAT-detected ones. In Fig.~\ref{CD}, the CD versus $\gamma$-ray luminosity of sources in our sample has been shown. The plot confirms that sources with large $\gamma$-ray luminosity also have high CD values. In the cases of J0754$+$3910 and J0903$+$4055, the upper limits on the $\gamma$-ray flux estimated over 8.5 years of LAT observations are compatible with the sensitivity limit estimated over 10 years of LAT observations, indicating that, accumulating more data, these sources may be detected by the LAT in the near future. 
 
\section{Summary}\label{summary}

With the advent of the {\em Fermi} satellite, our understanding of the radio and $\gamma$-ray connection in blazars significantly increased. BL Lacs are found to be the most abundant class of extragalactic objects observed by LAT. However, they are relatively weak radio sources and the parsec-scale structure has not been studied in detail for most of them. We have selected a sample of 42 BL Lacs included in the BZCAT catalogue with $z < 0.2$, located within the sky area covered by SDSS, and with no selection limit on $\gamma$-ray flux. Most of the sources (31/42) have a 1.4 GHz flux density that is lower than the sensitivity limit of VIPS (i.e. 85 mJy). These sources have been observed with the VLBA at 8 GHz and 15 GHz, and the results are reported in \citet{Liuzzo2013}. In this paper we have investigated the $\gamma$-ray properties of the radio fainter BL Lac sources and the connection with the radio properties.

Among the sources in the sample, 23 out of 42 have been detected by our LAT analysis, with an average photon index $\langle \Gamma \rangle = 1.87 \pm 0.06$ and a luminosity varying between 6$\times$10$^{43}$ and 6$\times$10$^{45}$ erg s$^{-1}$, a range of values typical of $\gamma$-ray
emitting BL Lacs and within the range of values observed for the other BL Lacs with $z < 0.2$ in the 3LAC. All sources reported in the 3FGL are confirmed by our analysis. In addition we have revealed three new sources with respect to the 3FGL. These three new detections have fluxes below $1.4\times10^{-8}$ \pflux and photon index 1.7--1.9. 

Based on the $\nu_{peak}^{syn}$, 30 sources in our sample are classified as HSP; in particular 19 out of 23 LAT-detected sources are HSP BL Lacs. An
anti-correlation between $\nu_{peak}^{syn}$ and $\gamma$-ray photon index is observed for the LAT-detected BL Lacs, in agreement with the blazar
sequence. Among the 23 LAT-detected sources, 19 are included in the 3FHL, for which a small difference between the photon index estimated in the 0.1--300 GeV and 10 GeV--2 TeV energy range has been observed. This suggests that the average $\gamma$-ray spectrum of these sources connects rather smoothly from 0.1 GeV to 2 TeV, as expected due to the low EBL attenuation at $z$ $<$ 0.2.
 
A compact radio component in the VLBA images has been detected in 22 out of the 23 LAT-detected sources, and only 6 out of 19 non-LAT-detected
sources. LAT-detected BL Lacs are more luminous on parsec scales with respect to non-LAT-detected sources and tend to have larger core dominance according to the Unified models. All sources with $P_{\rm\,core,\rm\,8 GHz}$ $>$ 2 $\times$10$^{24}$ W Hz$^{-1}$ have been detected in $\gamma$ rays by LAT, confirming the strong correlation between radio and $\gamma$-ray properties. In the same way, we have identified a possible population of
low-luminosity BL Lacs in both the radio and $\gamma$ rays, lacking a VLBI core, and with a small Doppler factor. This highlights an opportunity for future $\gamma$-ray observations to discover low-luminosity BL Lacs with {\em Fermi}-LAT and the next generation of $\gamma$-ray satellites.

Radio and $\gamma$-ray emission in blazars are both related to the presence of relativistic particles in jets. Among the 23 LAT-detected sources, 20 are classified as DD, indicating the presence of a relativistic jet in radio images. All these characteristics confirm that LAT-detected BL Lacs are dominated by Doppler boosting effects. However, three LAT-detected sources are classified as LD, with no evidence of relativistic jet with high Doppler factor in radio images and relatively low core dominance, showing non-classical properties for a $\gamma$-ray-emitting BL Lac. The properties of these peculiar sources will be investigated in detail by means of multi-frequency observations. 

\begin{acknowledgements}

We acknowledge financial contribution from grant PRIN$-$INAF$-$2011. This work was supported by the National Research Council of Science \& Technology (NST) granted by the International joint research project (EU-16-001). We thank the anonymous referee, M. Kadler, M. Orienti, A. Dominguez, R. Caputo, and D. Thompson for useful comments and suggestions. \\

The \textit{Fermi} LAT Collaboration acknowledges generous ongoing support from a number of agencies and institutes that have supported both the
development and the operation of the LAT as well as scientific data analysis. These include the National Aeronautics and Space Administration and the Department of Energy in the United States, the Commissariat \`a l'Energie Atomique and the Centre National de la Recherche Scientifique / Institut National de Physique Nucl\'eaire et de Physique des Particules in France, the Agenzia Spaziale Italiana and the Istituto Nazionale di Fisica Nucleare in Italy, the Ministry of Education, Culture, Sports, Science and Technology (MEXT), High Energy Accelerator Research Organization (KEK) and Japan Aerospace Exploration Agency (JAXA) in Japan, and the K.~A.~Wallenberg Foundation, the Swedish Research Council and the Swedish National Space Board in Sweden.
 
Additional support for science analysis during the operations phase is gratefully acknowledged from the Istituto Nazionale di Astrofisica in Italy and the Centre National d'\'Etudes Spatiales in France. This work performed in part under DOE Contract DE-AC02-76SF00515.

\end{acknowledgements}

\end{document}